\documentclass[prl,twocolumn,showpacs,floatfix]{revtex4-1}
\usepackage{amsmath}
\usepackage{amssymb}
\usepackage{graphicx}
\usepackage[english]{babel}
\usepackage{latexsym}
\usepackage{graphics}
\usepackage{subfigure}

\def\be{\begin{equation}}
\def\ee{\end{equation}}
\def\bea{\begin{eqnarray}}
\def\eea{\end{eqnarray}}
\def\bse{\begin{subequations}}
\def\ese{\end{subequations}}

\def\be{\begin{eqnarray}}
\def\ee{\end{eqnarray}}

\begin{document}

\title{Superconducting phase with a chiral $f$-wave pairing symmetry and
Majorana fermions induced in a hole-doped semiconductor}
\author{Li Mao$^{1}$, Junren Shi$^{2}$, Qian Niu$^{2,3}$}
\author{Chuanwei Zhang$^{1}$}
\thanks{cwzhang@wsu.edu}

\begin{abstract}
We show that a chiral $f+if$-wave superconducting pairing may be induced in
the lowest heavy hole band of a hole-doped semiconductor thin film through
proximity contact with an \textit{s}-wave superconductor. The chirality of
the pairing originates from the $3\pi $ Berry phase accumulated for a heavy
hole moving along a close path on the Fermi surface. There exist three
chiral gapless Majorana edge states, in consistence with the chiral $f+if$%
-wave pairing. We show the existence of zero energy Majorana fermions in
vortices in the semiconductor-superconductor heterostructure by solving the
Bogoliubov-de-Gennes equations numerically as well as analytically in the
strong confinement limit.
\end{abstract}

\affiliation{$^{1}$Department of Physics and Astronomy, Washington State University,
Pullman, Washington, 99164 USA\\
$^{2}$International Center for Quantum Materials, Peking University, Beijing
100871, China \\
$^{3}$Department of Physics, The University of Texas, Austin, Texas 78712
USA}
\pacs{74.20Rp, 03.65.Vf, 71.10.Pm, 74.45.+c}
\maketitle

Superconductors/superfluids with unconventional pairings have been an
important subject in condensed matter physics for many decades because of
their rich physics and important applications. There has been considerable
experimental evidence to support that the pairing symmetry in high-$T_{c}$
superconductors is the $d$-wave \cite{Lee}. The pairing symmetry in the
superfluid $^{3}$He systems has been found to be $p$-wave \cite{Osheroff}.
The superconducting order parameters in Sr$_{2}$RuO$_{4}$ and some
heavy-fermion materials are suggested to be the chiral $p_{x}+ip_{y}$ wave
\cite{Maeno}, although the true nature of the order parameters in these
materials has not been fully settled in experiments \cite{Xia,Moler}. The
importance of the chiral $p$-wave superconductor/superfluid is that the
quasiparticle excitation inside a vortex is a zero-energy Majorana fermion
with non-Abelian exchange statistics, which is a crucial ingredient for
topological quantum computation (TQC) \cite{Nayak}.

However, in contrast to the simple $s$-wave superconductor described by the
BCS theory, the theoretical description and experimental identification of
unconventional superconductivity and the associated exotic physics in
natural solid state systems are often difficult, and in many systems,
controversial. For instance, despite the tremendous technological potential,
the observation of the exotic properties such as quantum half-vortices and
non-Abelian statistics in Sr$_{2}$RuO$_{4}$ has been a serious problem
because of the small quasiparticle excitation energy gap as well as the
intrinsic spin-orbit coupling in the suggested $p$-wave order parameter \cite%
{Sarma}. Therefore it should be not only interesting, but also important to
investigate whether various unconventional superconducting pairings and the
associated exotic physics can be externally induced from conventional
\textit{s}-wave superconductors/superfluids \cite{Zhangcw,Fu,Sau,Alicea}.
For instance, schemes have been proposed recently to induce the zero-energy
Majorana fermions in the vortex core of conventional $s$-wave
superconductors that are proximately coupled to topological insulators or
electron-doped semiconductors \cite{Fu,Sau,Alicea}.

In this paper, we propose that a chiral $f+if$-wave superconducting pairing
can be induced in a \textit{hole-doped} semiconductor thin film through the
proximity contact with an \textit{s}-wave superconductor. To the best of our
knowledge, a chiral $f+if$-wave superconducting pairing and the associated
exotic physics have not been unequivocally identified in any condensed
matter system. The induced chiral $f+if$-wave pairing symmetry has a
topological origin: the geometric phase \cite{Pancharatnam,Berry1984} of
holes in the Bloch band. It is well known that an electron/hole evolving
adiabatically in the reciprocal space accumulates a geometric phase
associating with the adiabatic change of the quasi-momentum~\cite{Berry1984}%
, in analogy to the Aharanov-Bohm phase acquired by electron moving in the
real space in the presence of a magnetic field. The geometric phase is
nonzero in the hole-doped semiconductors with non-vanishing spin-orbit
coupling, which tunes an original \textit{s}-wave pairing into a chiral $%
f+if $-wave pairing for holes in the lowest energy band. The induced chiral $%
f+if$-wave superconductor has a full pairing gap in the 2D bulk, and $3$
gapless chiral Majorana fermions at the edge. By solving the
Bogoliubov-de-Gennes (BdG) equations analytically and numerically, we show
that there exists a Majorana zero energy state in the vortex core of the
semiconductor-superconductor heterostructure in some parameter regions. The
corresponding quasiparticle exchange statistics in this system is the same
as that for a chiral $p$-wave superconductor or superfluid, therefore the
proposed heterostructure can be used as a platform for observing non-Abelian
statistics and performing TQC. The advantage of using hole-doped, instead of
electron-doped, semiconductors for TQC is that hole-doped semiconductors
have stronger spin-orbit coupling due to the larger effective mass of holes
and the \textit{p}-like symmetry of the valence band, resulting in larger
carrier densities.

\begin{figure}[t]
\includegraphics[width=1\linewidth]{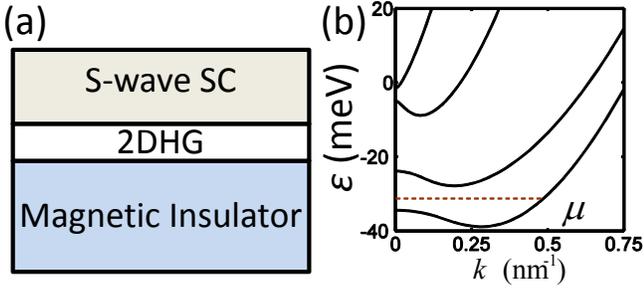} \vspace{-10pt}
\caption{(Color online) (a) A schematic illustration of the heterostructure
composed of a hole-doped semiconductor thin film, an \textit{s}-wave
superconductor, and a magnetic insulator. (b) The Band structure of the
hole-doped semiconductor. The lower (upper) two bands are the HH (LH),
respectively. $\protect\alpha =2\times 10^{5}$ m/s, $\hbar \protect\sqrt{%
\left\langle k_{z}^{2}\right\rangle }$ $=3\times 10^{-26}$ kg$\cdot $m/s, $%
\protect\gamma _{1}=6.92$, $\protect\gamma _{2}=2.1$, $h_{0}=1.75$ meV. The
parameters are chosen from GaAs (other materials such as InAs, InSb yield
the same physics). The hole density is $\sim 4\times 10^{12}$ cm$^{-2}$.}
\label{band}
\end{figure}

The physical system we consider is a heterostructure composed of an \textit{s%
}-wave superconductor, a hole-doped semiconductor thin film, and a magnetic
insulator (Fig. \ref{band}a). In the semiconductor thin film, the dynamics
of holes can be described by a single particle effective Hamiltonian that
contains both Luttinger four band model and spin-3/2 Rashba term \cite{Zhang}%
\begin{eqnarray}
H_{0} &=&\left[ \left( \gamma _{1}+5\gamma _{2}/2\right) \mathbf{k}%
^{2}-2\gamma _{2}\left( \mathbf{k}\cdot \mathbf{J}\right) ^{2}\right] /2m
\notag \\
&&+\alpha \left( \mathbf{J}\times \mathbf{k}\right) \cdot \hat{z}%
+2h_{0}J_{z}-\mu ,  \label{Ham1}
\end{eqnarray}%
where $\mathbf{J}$ is the total angular momentum operator for a spin-3/2
hole, $\gamma _{1}$ and $\gamma _{2}$ are the Luttinger parameters, $\mu $
is the chemical potential. Henceforth, we have set $\hbar =1$. The
confinement of the quantum well along the $\mathit{z}$ direction makes the
momentum be quantized on this axis, that is, $\left\langle
k_{z}\right\rangle \approx 0\,$, $\left\langle k_{z}^{2}\right\rangle
=\left( \pi /a\right) ^{2}$, where $a$ is the thickness of the quantum well.
$\alpha $ is the Rashba spin-orbit coupling strength. The crucial difference
between the Rashba terms in the 2D hole and electron gases is that $\mathbf{J%
}$ in the 2D hole gas is a spin-3/2 matrix, describing both the heavy holes
(HH) and light holes (LH). The term $2h_{0}J_{z}$ describes a Zeeman
splitting induced either through the polarization of the local magnetic
moments in the semiconductor \cite{Niu2} or the exchange field through the
contact with a magnetic insulator.

The eigenstates of the Hamiltonian (\ref{Ham1}) can be written as%
\begin{equation}
\Psi _{\mathbf{k}}=\left(
\begin{array}{cccc}
u_{0}e^{-i3\theta _{\mathbf{k}}}, & iu_{1}e^{-i2\theta _{\mathbf{k}}}, &
-u_{2}e^{-i\theta _{\mathbf{k}}}, & -iu_{3}%
\end{array}%
\right) ^{T}  \label{wavefunction}
\end{equation}%
where $\theta _{\mathbf{k}}$ is the azimuthal angle of $\mathbf{k}$, $u_{i}$
are functions of $k$ only and the eigenstates of the reduced Hamiltonian%
\begin{equation}
\bar{H}_{0}=-\gamma _{2}k^{2}J_{y}^{2}/m-\alpha kJ_{x}-\gamma
_{2}\left\langle k_{z}^{2}\right\rangle J_{z}^{2}/m+2h_{0}J_{z}.
\label{Ham2}
\end{equation}%
The particular choice of the wavefunction in (\ref{wavefunction}) ensures
that the wavefunction for the lowest HH band is single-valued at $\mathbf{k}%
=0$ (\textit{i.e.}, only $u_{3}\neq 0$ at $\mathbf{k}=0$). Additional total
phase $e^{im\theta _{\mathbf{k}}}$ need be multiplied to the wavefunctions
for the other bands to ensure the single value. In Fig. \ref{band}b, we plot
the eigenenergies with respect to $k$ for the 2D hole gas. The degeneracy
between different HH and LH bands at $k=0$ is lifted due to nonzero $%
\left\langle k_{z}^{2}\right\rangle $ and the Zeeman field $h_{0}$. In the
strong (weak) confinement region $2\gamma _{2}\left\langle
k_{z}^{2}\right\rangle /m>$ ($<$) $4h_{0}$, the second lowest energy band is
the HH (LH), and the corresponding wavefunction of the band can be written
as $\Psi _{\mathbf{k}}e^{i3\theta _{\mathbf{k}}}$ ($\Psi _{\mathbf{k}%
}e^{i\theta _{\mathbf{k}}}$).

\begin{figure}[t]
\includegraphics[width=1\linewidth]{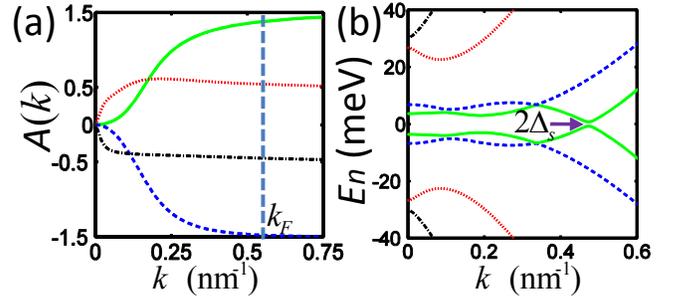} \vspace{-20pt}
\caption{(Color online) (a) Plot of $A(k)$ with respect to $k$. (b) Plot of
the bulk quasiparticle energy $E_{n}$ of the BdG equation (\protect\ref{BdG1}%
). Solid, dashed, dotted, and dashed dotted lines correspond to the lowest
to the highest bands in Fig. \protect\ref{band}b. $\protect\mu =-32.5$ eV.
The other parameters are the same as that in Fig. 1.}
\label{Berryphase}
\end{figure}

The proximity-induced superconductivity in the hole-doped semiconductor can
be described by the Hamiltonian \cite{book}
\begin{equation}
\hat{H}_{p}=\sum\nolimits_{m_{J}=1/2,3/2}\int d\mathbf{r}\left\{ \Delta
_{s}\left( \mathbf{r}\right) c_{m_{J}}^{\dag }c_{-m_{J}}^{\dag
}+H.c.\right\} ,  \label{pairing}
\end{equation}%
where $c_{m_{J}}^{\dag }$ are the creation operators for holes with the
angular momentum $m_{J}$ and $\Delta _{s}\left( \mathbf{r}\right) $ is the
proximity-induced gap. When the chemical potential lies between the lowest
two bands (Fig. \ref{band}b), and only the lowest HH band is occupied with
holes, the effective superconducting pairing for holes becomes%
\begin{equation}
\Delta _{eff}\propto \left\langle a_{\mathbf{k}}a_{-\mathbf{k}}\right\rangle
\propto i\Delta _{s}g\left( k\right) \exp \left( i3\theta _{\mathbf{k}%
}\right) ,  \label{order}
\end{equation}%
where $a_{\mathbf{k}}=\sum_{m_{J}}\chi _{m_{J}}c_{\mathbf{k}m_{J}}$ is the
annihilation operator for holes at the lowest HH state $\Psi _{\mathbf{k}}$,
with the coefficient $\chi _{m_{J}}=\Psi _{\mathbf{k}m_{J}}^{\ast }$, $%
g\left( k\right) =2\left( u_{0}u_{3}+u_{1}u_{2}\right) $. We\ have used $%
\Delta _{s}\propto \left\langle a_{\mathbf{k}m_{J}}a_{-\mathbf{k}\left(
-m_{J}\right) }\right\rangle $ and $\left\langle a_{\mathbf{k}m_{J}}a_{-%
\mathbf{k}m_{J}^{\prime }}\right\rangle =0$ if $m_{J}\neq -m_{J}^{\prime }$
to derive Eq. (\ref{order}). Clearly the pairing order $\Delta _{eff}$ has a
chiral $f+if$-wave symmetry. Around the Fermi surface $g\left( k\right)
\rightarrow 1$, that is, $\Delta _{eff}\rightarrow i\Delta _{s}\exp \left(
i3\theta _{\mathbf{k}}\right) $.

The $3\theta _{\mathbf{k}}$ phase in $\Delta _{eff}$ originates from a $3\pi
$ Berry phase accumulated when the holes move in the momentum space. In the
lowest HH band, the Berry phase for a hole along a loop on the Fermi surface
is
\begin{equation}
\phi =\int_{\mathbf{k}_{1}}^{\mathbf{k}_{2}}\mathbf{A\cdot dk}=A\left(
k_{F}\right) \delta \theta _{\mathbf{k}},  \label{phase}
\end{equation}%
where the Berry connection $\mathbf{A}=i\left\langle \Psi _{\mathbf{k}%
}\right\vert \nabla _{\mathbf{k}}\left\vert \Psi _{\mathbf{k}}\right\rangle
=A\left( k\right) \nabla \theta _{\mathbf{k}}$ with $A\left( k\right)
=\left( 3u_{0}^{2}+2u_{1}^{2}+u_{2}^{2}\right) $, $\delta \theta _{\mathbf{k}%
}=\theta _{\mathbf{k}_{2}}-\theta _{\mathbf{k}_{1}}$ is the change of the
azimuthal angle from $\mathbf{k}_{1}$ to $\mathbf{k}_{2}$. In Fig. \ref%
{Berryphase}a, we see $A\left( k\right) \rightarrow 3/2$ around the Fermi
surface for the lowest HH state, indicating a $3\delta \theta _{\mathbf{k}%
}/2 $ Berry phase ($3\pi $ for a close loop) for a single hole and $3\delta
\theta _{\mathbf{k}}$ phase for a Cooper pair. Therefore $\Delta _{eff}$ in
the lowest band has a phase factor $\exp \left( i3\theta _{\mathbf{k}%
}\right) $. Similarly, we can calculate the Berry connection for the other
three upper bands. In Fig. \ref{Berryphase}a, we see the corresponding Berry
phases are $-3\pi $, $\pi $, and $-\pi $, respectively, which means that the
superconducting pairing symmetries for holes at the upper HH, lower LH, and
upper LH bands are chiral $f-if$, $p_{x}+ip_{y}$, $p_{x}-ip_{y}$-waves,
respectively.

The physics origin of the chiral $f+if\,$-wave\ superconducting pairing in
the lowest HH band is more transparent in the strong confinement limit ($%
k\ll \sqrt{\left\langle k_{z}^{2}\right\rangle }$), where the four band
Hamiltonian (\ref{Ham1}) can be diagonalized into two effective two-band
Hamiltonians for the HH and LH respectively. The effective Hamiltonian $%
H_{hh}$ for the HH
\begin{equation}
H_{hh}=\eta _{0}k^{4}+\eta _{1}k^{2}+i\beta \left( k_{-}^{3}\sigma
_{+}-k_{+}^{3}\sigma _{-}\right) +3h_{0}\sigma _{z}-\bar{\mu}.  \label{Ham3}
\end{equation}%
Here $k_{\pm }=k_{x}\pm ik_{y}$, $\sigma _{\pm }=\left( \sigma _{x}\pm
i\sigma _{y}\right) /2$ are Pauli matrices applied on the two HH states
(denoted as pseudospin $\uparrow $ and $\downarrow $), $\eta _{i}$ are the
reduced coefficients, $\beta $ is the effective coupling strength, $\bar{\mu}
$ is the effective chemical potential. The Hamiltonian (\ref{Ham3}) is
similar as the Rashba type of Hamiltonian for electron-doped semiconductors
except that $k_{x}+ik_{y}$ is now replaced with $\left( k_{x}+ik_{y}\right)
^{3}=k\exp \left( i3\theta _{\mathbf{k}}\right) $ and there is a $k^{4}$
term to ensure the bands bend up for a large $k$. Therefore a chiral $f+if$%
-wave superconducting pairing is obtained when only the lowest HH band is
occupied with holes.

A chiral $f+if$-wave superconductor should have a full pairing gap in the 2D
bulk, and $\mathcal{C}=3$ gapless chiral Majorana fermions at the edge \cite%
{Wilczek}, where
\begin{equation}
\mathcal{C}=\frac{1}{2\pi }\sum\nolimits_{E_{n}<0}\int d^{2}\mathbf{k}\text{
}\mathbf{\Omega }_{z}^{n}  \label{Chernnumber}
\end{equation}%
is the first Chern number in the momentum space, $\mathbf{\Omega }_{z}^{n}=-2
$Im$\left\langle \partial \Phi _{n}/\partial k_{x}|\partial \Phi
_{n}/\partial k_{y}\right\rangle $ is the Berry curvature of the $n$-th
band. $E_{n}$ and $\Phi _{n}$ are eigenenergies and wavefunctions of the BdG
equation, which, in the Nambu spinor basis, can be written as%
\begin{equation}
\left(
\begin{array}{cc}
H_{0} & \Delta _{s}(\mathbf{r}) \\
\Delta _{s}^{\ast }(\mathbf{r}) & -\bar{\sigma}_{x}\tau _{y}H_{0}^{\ast
}\tau _{y}\bar{\sigma}_{x}%
\end{array}%
\right) \Phi _{n}(\mathbf{r})=E_{n}\Phi _{n}(\mathbf{r}).  \label{BdG1}
\end{equation}%
Here $\Phi _{n}(\mathbf{r})=[u_{n,3/2}$ $u_{n,1/2}$ $u_{n,-1/2}$ $u_{n,-3/2}$
$v_{n,-3/2}$ $v_{n,-1/2}$ $-v_{n,1/2}$ $-v_{n,3/2}]^{T}$ is the
quasiparticle wavefunction, $\bar{\sigma}_{x}=diag\left( \sigma _{x},\sigma
_{x}\right) $, $\tau _{y}=\left(
\begin{array}{cc}
0 & -iI_{2\times 2} \\
iI_{2\times 2} & 0%
\end{array}%
\right) $. In a uniform system with a constant $\Delta _{s}(\mathbf{r})$,
the BdG equation (\ref{BdG1}) can be solved in the momentum space and the
corresponding quasiparticle energy dispersions $E_{n}\left( k\right) $ are
plotted in Fig. \ref{Berryphase}b. We see a 2$\Delta _{s}$ energy gap is
opened at the Fermi surface. Using the eigenwavefunctions $\Phi _{n}$ in the
lowest four bands of the BdG equation (\ref{BdG1}) (i.e., $E_{n}<0$), we
confirm that the Chern number $\mathcal{C}=3$, which is consistent with the
chiral $f+if$-wave superconducting pairing and yields 3 gapless chiral
Majorana fermions at the edge of the superconductor.

\begin{figure}[t]
\includegraphics[width=0.7\linewidth]{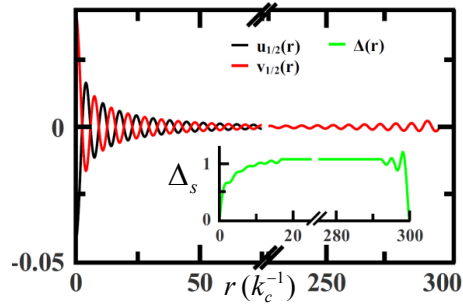} \vspace{-10pt}
\caption{(Color online) Plot of the wavefunction of the zero energy state. $%
\protect\mu =-32.5$ meV lies in the gap between two HH bands. The other
parameters are the same as that in Fig. 1. Inset: Plot of the \textit{s}%
-wave pairing gap with a vortex.}
\label{zerostate}
\end{figure}

The chiral $f+if$-wave pairing may lead to novel exotic physics that has not
been explored before (e.g., fractional Josephson effects \cite{Kwon}). Here
we focus on the Majorana fermions in vortices in the heterostructure that
can be used for TQC. In the presence of a vortex in the heterostructure, the
pairing order parameter takes the form $\Delta _{s}(\mathbf{r})=\Delta
_{s}(r)e^{i\theta }$. For simplicity of the calculation, we consider a 2D
cylinder geometry with a hard wall at the radius $r=R$ and a single vortex
at $r=0$. This system preserves the rotation symmetry and the BdG equation
can be decoupled into different angular momentum channels indexed by $l$
with the corresponding spinor wavefunction $\Phi _{n}^{l}(\mathbf{r}%
)=e^{il\theta }[u_{n,3/2}^{l}e^{-i\theta }$,$u_{n,1/2}^{l}$,$%
u_{n,-1/2}^{l}e^{i\theta }$,$u_{n,-3/2}^{l}e^{2i\theta }$, $%
v_{n,-3/2}^{l}e^{-2i\theta }$,$v_{n,-1/2}^{l}e^{-i\theta }$,$v_{n,1/2}^{l}$,$%
v_{n,3/2}^{l}e^{i\theta }]^{T}$. Here $u$ and $v$ are functions of $r$ only.
The special form of $\Phi _{n}^{l}(\mathbf{r})$ is chosen to preserve the
particle-hole symmetry at $l=0$ and to remove the $\theta $ dependence in
the BdG equation (\ref{BdG1}). If $\Phi _{n}^{l}(\mathbf{r})$ is a solution
with an energy $E$, then there is another solution with the energy $-E$ in
the $-l$ channel. Henceforth we only consider $E\geq 0$ solutions.

Generally the BdG equation (\ref{BdG1})\ with a vortex cannot be solved
analytically. Here we numerically solve the Eq. (\ref{BdG1})\ and calculate
the quasiparticle eigenenergies and eigenwavefunctions. In the calculation,
we use the pairing gap $\Delta _{s}$ from a self-consistence solution of the
BdG equation for a pure $s$-wave superconductor with a small size of the
system $R=25k_{c}^{-1}$ (the Fermi vector $k_{c}$ for the \textit{s}-wave
superconductor is chosen as 0.5 nm$^{-1}$). Because the pairing gap
approaches the bulk value in a distance much larger than $k_{c}^{-1}$, we
can extend the pairing gap to a larger system $R=300k_{c}^{-1}$ by inserting
the uniform bulk value (see the inset in Fig. \ref{zerostate}). We find that
there exists a unique zero energy solution when the chemical potential $\mu $
lies in the gap between the lowest two HH bands (Fig. \ref{band}b). In Fig. %
\ref{zerostate}, we plot the two components $u_{0,1/2}^{0}(r)$ and $%
v_{0,1/2}^{0}(r)$ of the zero energy wavefunction $\Phi _{0}^{0}(\mathbf{r})$
and find $u_{0,1/2}^{0}(r)=-v_{0,1/2}^{0}(r)$. We also confirm that $%
u_{0,m_{J}}^{0}(r)=-v_{0,m_{J}}^{0}(r)$ for other $m_{J}$. Therefore the
Bogoliubov quasiparticle operator, defined as
\begin{equation}
\gamma _{n}^{\dag }=i\int d\mathbf{r}\sum\nolimits_{m_{J}}\left[ u_{nm_{J}}(%
\mathbf{r})c_{m_{J}}^{\dag }\left( \mathbf{r}\right) +v_{nm_{J}}(\mathbf{r}%
)c_{m_{J}}\left( \mathbf{r}\right) \right] ,  \label{Bog}
\end{equation}%
satisfies $\gamma _{0}^{\dag }=\gamma _{0}$, which is a self-Hermitian
Majorana operator. Consider two Majorana operators $\gamma _{A}$ and $\gamma
_{B}$ in two vortices. It is easy to show $\gamma _{A}\rightarrow \gamma
_{B} $, $\gamma _{B}\rightarrow -\gamma _{A}$ upon an exchange of two
vortices \cite{Stone}. Therefore the Majorana zero energy modes satisfy the
same non-Abelian braiding statistics as that in a chiral \textit{p}-wave
superconductor/superfluid \cite{Ivanov,Tewari} and can be used for TQC.

\begin{figure}[t]
\includegraphics[width=0.7\linewidth]{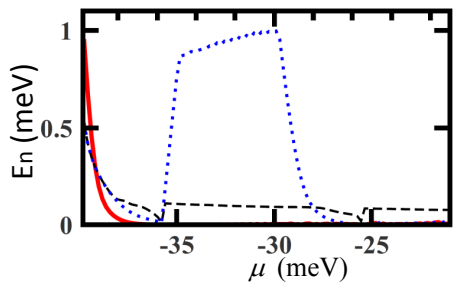} \vspace{-10pt}
\caption{Plot of the quasiparticle energies in a vortex core with respect to
the chemical potential $\protect\mu $. Solid: zero energy state; Dashed: the
minigap. Dotted: the bulk excitation gap. The parameters are the same as
that in Fig. 1.}
\label{spectrum}
\end{figure}

In Fig. \ref{spectrum}, we plot the zero energy state (the lowest energy
level at the $l=0$ channel), the bulk excitation gap (the first excitation
at the $l=0$ channel), and the minigap energy (the lowest energy level in
the vortex core at the $l=1$ channel) with respect to the chemical potential
$\mu $. When $\mu $ lies in the gap between the lowest two HH bands, there
exists a unique zero energy solution, which originates from the broken time
reversal symmetry of the chiral $f+if$-wave superconducting pairing. The
minigap is the topological gap protecting the Majorana fermions at the zero
energy states and the associated non-Abelian braiding statistics from finite
temperature effects. The numerical results show that the magnitude of the
minigap is at the order between $\Delta _{s}$ and $\Delta _{s}^{2}/E_{F}$.
Additional numerical calculation shows that Majorana zero energy solutions
also exist for a vortex with a winding number $-3$ or other odd numbers. We
also find that the Majorana zero energy modes exist when the second lowest
band is LH, instead of HH.

The existence of the Majorana zero energy modes can also be demonstrated
analytically in the strong confinement limit. In this limit, the single
particle Hamiltonian $H_{0}$ is replaced with $H_{hh}$ in (\ref{Ham3}). The
spinor wavefunction at the angular momentum $l$ channel changes to $\Phi
_{n}^{l}(\mathbf{r})=e^{il\theta }[u_{n\uparrow }^{l}e^{-i\theta }$,$%
u_{n\downarrow }^{l}e^{2i\theta }$,$v_{n\downarrow }^{l}e^{-2i\theta }$,$%
-v_{n\uparrow }^{l}e^{i\theta }]^{T}$. The corresponding BdG equation can be
further reduced to a $2\times 2$ matrix form
\begin{equation}
\left(
\begin{array}{cc}
\digamma _{0}-\bar{\mu}+3h_{0} & \beta \digamma _{1}+\lambda \Delta _{s} \\
-\beta \digamma _{2}-\lambda \Delta _{s} & \digamma _{4}-\bar{\mu}-3h_{0}%
\end{array}%
\right) \left(
\begin{array}{c}
u_{\uparrow }\left( r\right) \\
u_{\downarrow }\left( r\right)%
\end{array}%
\right) =0  \label{BdG3}
\end{equation}%
for a zero energy state after $\theta $ is eliminated using the wavefunction
$\Phi _{n}^{l}(\mathbf{r})$ and the particle-hole symmetry of the
wavefunction is taken into account \cite{Sau}. Here $\digamma _{0}\left(
r\right) =\eta _{0}\left( Q-r^{-2}\right) ^{2}+\eta _{1}\left(
Q-r^{-2}\right) $, $\digamma _{1}\left( r\right) =\partial _{r}\left(
\partial _{r}+r^{-1}\right) \left( \partial _{r}+2r^{-1}\right) $, $\digamma
_{2}\left( r\right) =\left( \partial _{r}+r^{-1}\right) \partial _{r}\left(
\partial _{r}-r^{-1}\right) $, $\digamma _{4}\left( r\right) =\eta
_{0}\left( Q-4r^{-2}\right) ^{2}+\eta _{1}\left( Q-4r^{-2}\right) $, $Q=%
\frac{\partial ^{2}}{\partial r^{2}}+\frac{1}{r}\frac{\partial }{\partial r}$%
, $u_{\sigma }\left( r\right) =\lambda v_{\sigma }\left( r\right) $. We
approximate the radial dependence of $\Delta _{s}$ as a step function,
\textit{i.e.}, $\Delta _{s}=0$ for $0\leq r\leq \xi $, and $\Delta _{0}$ for
$r>\xi $. Detailed analysis of the wavefunction $\left( u_{\uparrow }\left(
r\right) ,u_{\downarrow }\left( r\right) \right) ^{T}$ shows that there are
four and five independent solutions of Eq. (\ref{BdG3}) inside and outside
the vortex core respectively in the parameter region $\lambda =-1$ and $\bar{%
\mu}^{2}+\Delta ^{2}<9h_{0}^{2}$. The corresponding 9 unknown superposition
coefficients for the total wavefunction match with the 9 constraints from
the continuity of the wavefunction (up to the third order derivative) and
the normalization condition, yielding a unique zero energy solution.

In summary, we show that a chiral $f+if$-wave superconducting pairing and
the associate Majorana physics may be induced in a hole-doped semiconductor
thin film through the proximity contact with an \textit{s}-wave
superconductor. The proposed Berry phase mechanism in this system presents a
new possibility for studying unconventional pairing symmetry, which is
distinctly different from the conventional scenario in which the pairing is
induced by the Boson-exchange electron-electron interaction mechanism.

\begin{acknowledgments}
C.Z. thanks Kai Sun and Jay Sau for valuable discussion. L.M. and C.Z are supported by DARPA-MTO (FA955-10-1-0497), DARPA-YFA (N66001-10-1-4025)
and ARO (W911NF-09-1-0248). J.S. is supported by 973 program No. 2009CB929101 and NSFC 10734110. Q.N. is supported in part by DoE
(DE-FG02-02ER45958, Division of Materials Sciences and Engineering) and the
Welch Foundation (F-1255).
\end{acknowledgments}

\end{document}